\documentclass[preprint,12pt]{elsarticle}

\usepackage{amssymb}
\usepackage{graphicx}
\graphicspath{ {./images/} }

\journal{arXiv}

\begin{document}

\begin{frontmatter}



\title{Development of a neutrino detector capable of operating in space}


\author[inst1]{Nickolas Solomey}

\affiliation[inst1]{organization={Physics Division, Wichita State University},%Department and Organization
            addressline={1845 Fairmount St.}, 
            city={Wichita},
            postcode={67260}, 
            state={Kansas},
            country={USA}}

\begin{abstract}
The $\nu$SOL experiment to operate a neutrino detector close to the Sun is building a small test detector to orbit the
 Earth to test the concept in space. This detector concept is to provide a new way to detect neutrinos unshielded 
in space. A double delayed coincidence on Gallium nuclei that have a large cross section for solar neutrino interactions emitting a conversion electron and converting the nuclei into an excited state of Germanium, which decays with a well-known energy and half-life. This unique signature permits operation of the detector volume mostly unshielded in space with a high single particle counting rate from gamma and cosmic ray events. The test detector concept which has been studied in the lab and is planned for a year of operations orbiting Earth which is scheduled for launch in late 2024. It will be surrounded by an 
active veto and shielding will be operated in a polar orbit around the Earth to validate the detector concept and 
study detailed background spectrums that can fake the double timing and energy signature from random galactic cosmic or gamma rays. The success of this new technology development will permit the design of a larger spacecraft with a mission to fly close to the Sun and is of importance to the primary science mission of the Heliophysics division of NASA Space Science Mission Directorate, which is to better understand the Sun by measuring details of our Sun's fusion
 core.

\end{abstract}

%%\includegraphics{images/shieldveto.png}

\begin{highlights}
\item Neutrino Physics in Space
\item Dark Matter search in Space
\item Presented at the 15th Pisa meeting on Advanced Detector, June 2022 Elba, Italy
\end{highlights}

\begin{keyword}
solar physics \sep neutrino detector \sep dark matter
\PACS 26.65
\MSC 11956
\end{keyword}

\end{frontmatter}



\section{Introduction}
Neutrino detectors are normally large and deep underground, needing as much mass as possible to capture more neutrinos and underground to reduce backgrounds. If a neutrino detector could operate in space, what advantages could it bring to perform unique science not possible on Earth? A mobile neutrino laboratory would allow moving closer to
 the Sun where the neutrino flux is much larger, probing the processes at the core of the Sun. Moving away reduces
 the Solar neutrino background, revealing other processes such as Dark Matter or Galactic neutrinos currently hidden by a large Solar neutrino presence at the Earth.

This project has three distinct possibilities for science:
\begin{enumerate}
    \item The 1/r$^2$ neutrino flux closer to the Sun (see Table 1) would provide 1000× larger neutrino rates at 7 solar radii distance where the current NASA Parker Solar probe operates and 10,000× more neutrino rate at 3 solar radii where some NASA scientists think it is possible to go. New science would be a large-number statistics of solar 
neutrino emission, but by being 30x closer to the Sun this would permit the study of the size of the fusion emission region, and fundamental physics looking for the transition from coherent to de-coherent neutrino oscillations. 
Also, the presence of Dark Matter gravitationally trapped in Sun’s core could be indirectly detected by observing
the displaced solar fusion region.
    \item Because the neutrino has a non-zero mass, known from the existence of neutrino oscillations, the Sun itself bending space creates a gravitational focus [1]. For neutrinos, this gravitational focus would be very close to
us at 20 to 40 AU [2]: a distance easily reachable with current technology, unlike the gravitational focus for light at 450 to 750 AU. This solar neutrino focus could be a testing ground for gravitational lens experiments [3,4].
 The galactic core, 25,000 light years away from us and twice the apparent size of the moon, is the 2nd largest neutrino source in the sky after the Sun. The galactic core not only has many neutrino-producing stars, but also ~10,000 neutron or black holes in the central cubic parsec. Matter falling into these objects is crushed, producing neutrons and emitting neutrinos of higher energy than solar fusion neutrinos. A detector at the solar neutrino gravitational lens focus would permit imaging of the galactic core. Just finding the neutrino gravitational focus of the Sun would be a new way to measure the neutrino mass.
    \item Deviations from the expected 1/r$^2$ solar neutrino curve are an indication of new science. A deviation closer to the Sun observes changes due solely to the distance from a non-point source of solar neutrinos inside the Sun. A deviation away from the Sun is the direct observation of Dark Matter or galactic neutrinos. Under previous phases of NIAC [5,6], we devised a way to place a neutrino detector in space. Operation and detection in space is a 
game changer, enabling new measurements for Astrophysics, Heliophysics, and elementary particle physics, all of great importance.
\end{enumerate}

%%% Table of Distances
\begin{table}[htbp]
    \centering
    \begin{tabular}{|l|rl|}
        \textbf{Distance from Sun} & \textbf{Flux relati}&\hspace{-4.2mm}\textbf{ve to Earth} \\\hline\hline
        696342 km ($R_\odot$)& 46200&\\\hline
        1500000 km ($\sim3R_\odot$) & 10000&\\\hline
        4700000 km ($\sim7R_\odot$) & 1000&\\\hline
        15000000 km & 100&\\\hline
        47434000 km & 10&\\\hline
        Mercury & 6&\hspace{-4.2mm}.4\\\hline
        Venus & 1&\hspace{-4.2mm}.9\\\hline
        Earth & 1&\hspace{-4.2mm}.0\\\hline
        Mars & 0&\hspace{-4.2mm}.4\\\hline
        Asteroid Belt & 0&\hspace{-4.2mm}.1\\\hline
        Jupiter & 0&\hspace{-4.2mm}.037\\\hline
        Saturn & 0&\hspace{-4.2mm}.011\\\hline
        Uranus & 0&\hspace{-4.2mm}.0027\\\hline
        Neptune & 0&\hspace{-4.2mm}.00111\\\hline
        Pluto & 0&\hspace{-4.2mm}.00064\\\hline
        KSP & 0&\hspace{-4.2mm}.0002\\\hline
        Voyager 1 (2015) & 0&\hspace{-4.2mm}.00006\\\hline
    \end{tabular}
    \caption{Intensity of solar neutrinos at various distances from the sun.}
    \label{tab:fluxTable}
\end{table}
%%% End Table of Distances

This first idea would be of the most interest to the NASA Heliophysics division. We have under the NIAC Phase-2 grant done a detailed study of orbits close to the Sun and neutrino rates at various energies from various fusion processes. In one such mission design leaving Earth using a Falcon Heavy rocket and using Venus as a sling shot with every other 90-day orbit bringing the spacecraft closer and closer to the Sun on successive orbit 
 can produce a large amount, 100 per year, of neutrino interactions in the detector. The cost of such a mission 
 seems to fit into Small Explorer class spacecraft limitations and within 5 years of operations for a Technology Demonstrator flight that would do some new, unique science. These fusion neutrinos are only interactions not those 
that can be observed and reconstructed for physics analysis. That number would be reduced by 33\%, since two-thirds of the neutrino interactions on Ga produce an excited state nuclear resonance needed for the double time delayed
 coincidence and those events need to be further studied for detection efficiencies, but this still shows a substantial number of observable events even after these reductions.

The other two science missions possible are a dark matter explorer and a solar gravitation lens explorer. A neutrino gravitational focus would dramatically increase neutrino intensity due to the large light collecting power of a
 gravitational lens and the larger number of stars producing fusion neutrinos in the Galactic core or plane of the
 Galaxy. Both of these science missions would be of interest to the NASA Astrophysics division. Our preliminary study started with the assumption that the number of neutrinos produced by any star is directly proportional to the brightness of the star, and inverse square law of the distance with respect to the lens formed by the gravitational focus and its “light” collecting power. The estimate was made by integrating the 1991 Kent et al. [8] model of the luminosity distribution of the Milky Way galactic core. The size of the stellar neutrino signal at the Neutrino Gravitational Focus is also dependent on the field of view of the neutrino detector, which is a function of the diameter and depth of the neutrino detector. This resulted with the assumption of a 1 m diameter and 20 cm thick neutrino detector detecting 880× to 8800× more neutrinos from the galactic core at the Neutrino Gravitational Focus than from 
solar neutrinos at Earth directly from our Sun. We have simulated a space-craft flight and this science looks promising.

\begin{figure}[h]
\caption{GAGG Crystal sensor for the space flight test of the neutrino detector, each quadrant is read out by four SiPM photo sensors.}
\centering
\includegraphics[width=0.48\textwidth]{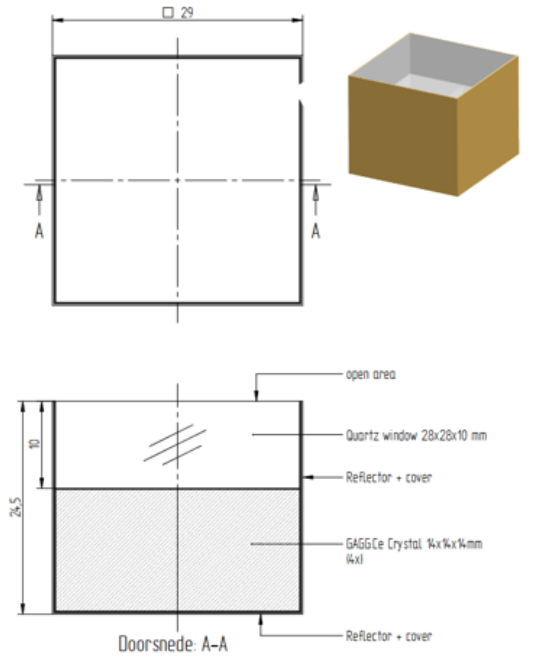}
\end{figure}
\begin{figure}[h]
\caption{Detector response with a $^{90}$Sr source, which should be similar to a neutrino conversion electron signal.}
\centering
\includegraphics[width=0.48\textwidth]{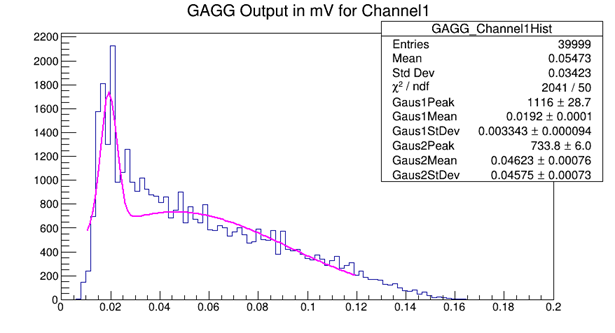}
\end{figure}
\begin{figure}[h]
\caption{Detector response with a $^{57}$Co source, which should be similar to the secondary gamma emission signal, the right peak is the 122 keV gamma signal and the left smaller peak the 46 keV escape peak.}
\centering
\includegraphics[width=0.48\textwidth]{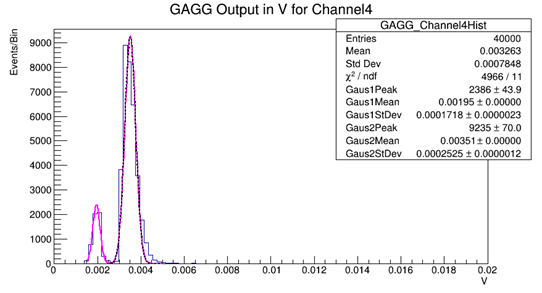}
\end{figure}

\section{Detector Technical Approach}
Observing neutrino interactions in space does not only need an increased flux of neutrinos but needs a technique to separate neutrino interactions from among the multiple cosmic and gamma ray background interactions. We use a time delayed double coincidence, similar to a techniques first used by Cohen and Reines [7], along with shielding 
from lower energy X-rays in space but designed for solar neutrinos. Our NIAC Phase-1 grant studied this idea, using Gallium to convert a neutrino into an electron and an excited state of Germanium 60\% of the time. The excited state has a characteristic half-life producing a delayed secondary pulse of a well-defined Gamma ray (see Table 2). Using a search for such a process eliminates most random background events. An expected 60\% conversion efficiency is high and does not degrade the performance. The increased flux of solar neutrinos close to the Sun allows us to accept the 40\% loss, but there is also a penalty of only getting slightly higher neutrino energy since this process has a threshold of 0.405 MeV. Our detector for the space flight test is a GAGG crystal that is 20\% Gallium, with fast 80 ns timing and high light yield of 50,000 photons per MeV of deposited energy, and is shown schematically in Figure 1.

%%% Table of Energies/times
\begin{table}[htp!]
\begin{center}
\begin{tabular}{ | c | c | c | c | }
\hline
{Reaction} & {Threshold} & {Photon Energy} & {Half Life} \\
 \hline\hline
 Ge$_{32}^{71E2}+e^-$& 0.408 MeV & 0.175 MeV & 79 ns \\  
 \hline
 Ge$_{32}^{69M1}+e^-$& 2.313 MeV & 0.086 MeV & 5 $\mu$s\\
 \hline Ge$_{32}^{69M2}+e^-$&2.624 MeV &0.397 MeV &2.8 $\mu$s\\
 \hline
\end{tabular}
\caption[Table of the possible reactions of neutrinos with gallium and their decay products.]{Table of the possible reactions of neutrinos with gallium and their decay products.}
\label{tab:NeutrinoReactions}
\end{center}
\end{table}
%%% End Table of Energies/times

Each of the four GAGG cubes is 14x14x14 mm$^3$ each read out by 4 SiPM photo sensor. A Beta source of $^{90}$Sr was used to characterize the performance of the conversion electron shown in Figure 2 and multiple gamma sources such as $^{57}$Co, $^{133}$Ba and $^{137}$Cs where used to characterize the performance of the nuclear excited state secondary emission decay comparable of the $^{71}$Ge or $^{69}$Ge emission, see Figure 3. The linearity and energy resolution of the GAGG detector for these lab tests are shown in figure 4 and 5 and although a GAGG crystal with a single photon counting sensor should be 4\% energy resolution due to using multiple SiPM these are slightly worse but are expected to be sufficient for our space flight test. 

The GAGG sensor will be encased in a plastic scintillator veto which is read out by a single photon sensitive photo-tube, and that whole assemble will be encased in a passive shield, see figure 6. The passive shield cannot be pure metal because all components in the CubeSat must be assured to burn up upon re-entry into the Earth atmosphere, so for this test we are using a mixture of Tungsten dust in epoxy that make the equivalent density as iron. All of this fits nicely into a 3U CubeSat. MSFC electronics group is building three dedicated flight cards for high voltage, discriminator threshold and FPGA logic.

\begin{figure}[h]
\caption{Energy linearity response of detector four quadrants (one is high light yield) using various gamma source.}
\centering
\includegraphics[width=0.48\textwidth]{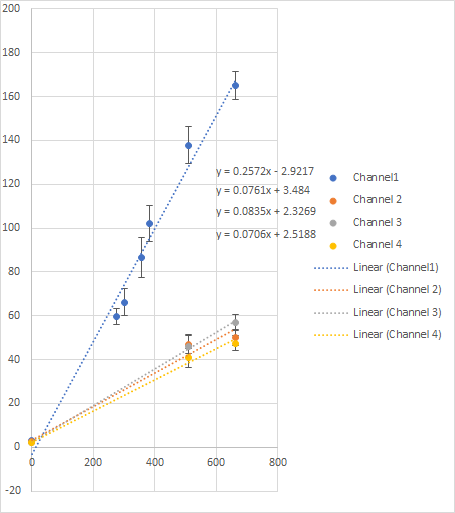}
\end{figure}
\begin{figure}[h]
\caption{Energy resolution far various gamma sources of different energy.}
\centering
\includegraphics[width=0.48\textwidth]{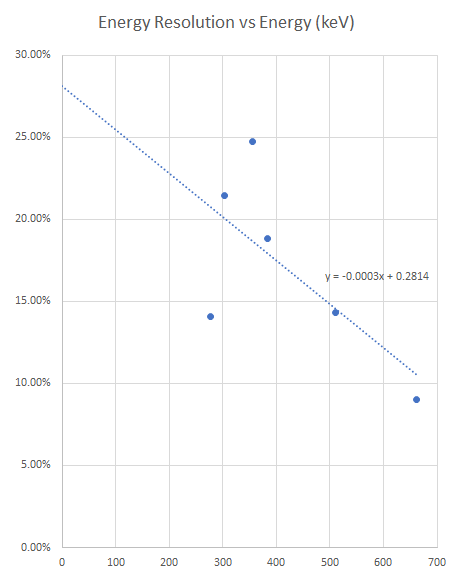}
\end{figure}
\begin{figure}[h]
\caption{GAGG detector sensor inside a veto and encased in a passive shield that fits into a 3U CubeSat.}
\centering
\includegraphics[width=0.48\textwidth]{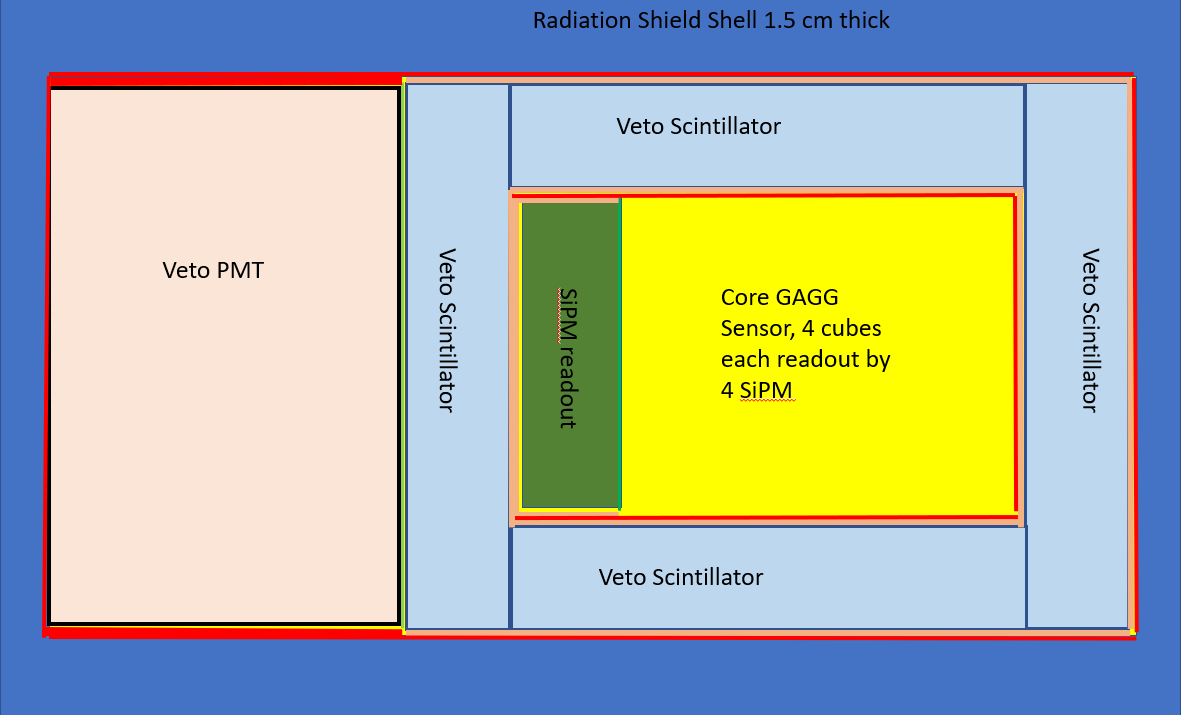}
\end{figure}

\section{Conclusion}
The main goal of this project is development and testing of a neutrino detector capable of operating in space, for more details see our project NIM-A full article submission [9]. This cube-sat in Earth orbit will not detect any neutrinos, but rather prove the detector is capable of space operations and to measure true background in space outside of the Earth’s radiation belts. Data from this experiment will
 be used to validate simulated detector and veto performance. The success of this project will be a true game changer to science for NASA and, all-be-it challenging, this is a necessary step along the development path to a possible missions with a neutrino detector capable of operating in deep space. The lab performance of the GAGG detector
 crystal for both gamma and beta sources is in the region of timing and light yield to permit the double timing delayed coincidence but space flight tests are expected from its operation in space. We look forward to the detector 
launch in 2024 and subsequent data analysis of its performance for this new science application.

\section*{Acknowledgments}
I wish to thank all of the students who have worked on this project at Wichita State University: C. Gimar, A. Nelson, L. Buchele, T. Nolan, S. Richardson, B. Doty, J. Novak, K. Messick and J. Folkerts. In addition to the other faculty who
have helped such a H. Meyer, H. Kwon and A. Dutta. Outside of Wichita State this project has also involved M. Christl and N. Barghouty of MSFC, B. Sutton of JPL and R. McTaggart of South Dakota State University, without who's help we would not have developed this project as fast as we did. This work was supported by the NASA Innovation Advanced Concept program in the Space Technology Mission Directorate, and I greatly appreciate their support. Further support was provided by a gift from Bill Simon to the Wichita State University Foundation for use on this project, and Fermi National Accelerator Lab has generously loaned us scintillator paddles, NIM and Camac electronics equipment for a cosmic ray test stand.

\section*{References}
[1] A. Einstein, Lens-Like Action of a Star by the Deviation of Light in the Gravitational Field, Science, Vol. 84
 (1936), p. 506.

[2] Y. Demkov and A. Puchkov, Gravitational focusing of cosmic neutrinos by the solar interior,
Phys. Rev. D v61, 083001.

[3] G.A. Landis, Mission to the Gravitational Focus of the Sun: A Critical Analysis, 23 April 2016, arXiv:1604.063
51v2 [astro-ph.EP].

[4] S.G. Turyshev and B-G. Andersson, “The 550-AU Mission: a critical discussion”, Mon. Not. R. Astron. Soc. 341, 
pp. 577-582 (2003).

[5] N. Solomey (PI), NASA Innovation and Advanced Concept 2018 Grant “Astrophysics and Technical Study of a Solar 
Neutrino Spacecraft”, May 15, 2018 to Feb. 14, 2019, and N. Solomey, Studying the Sun's Nuclear Furnace with a Neutrino Detector Spacecraft in Close Solar Orbit, AAS/Solar Physics Division, Abstracts\# 47 Presentation and poster 
P7-26, Boulder Colorado, June 2016.

[6] N. Solomey, M. Christl, C. Gimar, A. Nelsen, R. McTaggart and H. Meyer, Astrophysics, Technical and Mission Study of a Solar Neutrino Spacecraft, 27 Sept. 2018, NASA NIAC symposium, Boston, MA, see the online video link: https://livestream.com/viewnow/NIAC2018/videos/180892751.

[7] C. L Cowan Jr.; F. Reines; F. B. Harrison; H. W. Kruse; A. D McGuire (July 20, 1956). ``Detection of the Free 
Neutrino: a Confirmation''. Science. 124 (3212): 103–4.

[8] Kent, S.M., Dame, T.M., Fazio, G., 1991, Galactic structure from the Spacelab infrared telescope. II - Luminosity models of the Milky Way, ApJ 378, 131.

[9] Solomey, N. et al., Design of a Space-based Near Solar Neutrino Detector for the $\nu$SOL Experiment, submitted June 2022 to NIM-A, see arXiv.org article https://arxiv.org/abs/2206.00703.

%% \bibliographystyle{elsarticle-num} 
%% \bibliography{cas-refs}





% \end{thebibliography}
\end{document}